\documentclass[pra,twocolumn,showpacs,amsmath,amssymb,amsfonts]{revtex4}
\usepackage{bm}
\usepackage{graphicx}

\newcommand{\be}{\begin{equation}}
\newcommand{\ee}{\end{equation}}
\newcommand{\bea}{\begin{eqnarray}}
\newcommand{\eea}{\end{eqnarray}}
\newcommand{\beann}{\begin{eqnarray*}}
\newcommand{\eeann}{\end{eqnarray*}}
\newcommand{\besa}[1]{\begin{subequations}\label{#1}\begin{eqnarray}}
\newcommand{\eesa}{\end{eqnarray}\end{subequations}}

\newcommand{\ket}[1]{\, | #1 \rangle}

\newcommand{\hlf}{\mbox{$\frac{1}{2}$}}
\newcommand{\la}{\lambda}
\newcommand{\La}{\Lambda}
\newcommand{\om}{\omega}
\newcommand{\Om}{\Omega}
\newcommand{\ga}{\gamma}

\newcommand{\de}{\delta}
\newcommand{\De}{\Delta}
\newcommand{\ka}{\kappa}
\newcommand{\eps}{\epsilon}
\newcommand{\br}{\mathbf{r}}
\newcommand{\brp}{\mathbf{r}^{\prime}}
\newcommand{\bk}{\mathbf{k}}
\newcommand{\Ee}{\mathcal{E}}
\newcommand{\Pe}{\mathcal{P}}
\newcommand{\Ie}{\mathcal{I}}
\newcommand{\psih}{\hat{\psi}}
\newcommand{\gh}{\hat{g}}
\newcommand{\eh}{\hat{e}}
\newcommand{\sh}{\hat{s}}

\newcommand{\Fh}{\hat{\cal F}}

\newcommand{\lra}{\leftrightarrow}
\newcommand{\ddt}{\frac{\partial}{\partial t}}

\begin{document}

\title{Tunable photonic band gaps with 
coherently driven atoms in optical lattices}

\author{David Petrosyan}
\email[E-mail: ]{dap@iesl.forth.gr}
\affiliation{Institute of Electronic Structure \& Laser, 
FORTH, 71110 Heraklion, Crete, Greece}

\date{\today}

\begin{abstract}
Optical lattice loaded with cold atoms can exhibit a tunable photonic 
band gap for a weak probe field under the conditions of electromagnetically
induced transparency. This system possesses a number of advantageous 
properties, including reduced relaxation of Raman coherence and
the associated probe absorption, and simultaneous enhancement of 
the index modulation and the resulting reflectivity of the medium.
This flexible system has a potential to serve as a testbed of various 
designs for the linear and nonlinear photonic band gap materials at a 
very low light level and can be employed for realizing deterministic 
entanglement between weak quantum fields
\end{abstract}


\maketitle

\section{Introduction}

The properties of waves in spatially periodic media are studied
in various branches of physics, including acoustics, electromagnetism 
and quantum mechanics, initially called wave mechanics. Some of the 
fundamental properties of solid-state systems stem from the interplay 
of wave-like behavior of amplitudes describing the mobile electrons and
the spatially periodic potential created by the crystal lattice, which 
can result in the forbidden energy bands, or gaps, for the electrons 
\cite{SolStPh}. Analogous effects exist for electromagnetic waves in 
photonic crystal structures, where the refractive index is a periodic
function of spatial coordinates, resulting in the photonic band gaps 
(PBGs) \cite{pc-books,yariv}. 

Recently, spectacular progress has been achieved in cooling and trapping 
atoms in the optical lattices (OLs)---spatially periodic dipole potentials
induced by off resonant laser fields \cite{OptLatRev}. The relevant 
parameters of these systems can be controlled with very high precision 
and can be tuned to implement with unprecedented accuracy some of the 
fundamental models of condensed matter physics. A particularly relevant 
for the present studies achievement has been the demonstration of the 
transition from the superfluid to the Mott insulator phase with a 
commensurate number of bosonic atoms per lattice site \cite{mottOL}. 

As discussed below, OLs loaded with cold atoms interacting with a weak 
probe field and simultaneously driven by a strong coherent field can serve 
as a convenient tunable platform for the simulation and studies of light 
transmission and Bragg reflection in periodic media exhibiting PBGs. 
Earlier studies of related systems include a weak Bragg reflection of the 
probe field from the sparsely occupied by atoms 1D OL \cite{Phillips}, 
while more recently, optically induced 1D PBGs in uniform (gaseous) 
atomic media were realized \cite{swStEITPBG,swDrEITPBG} employing the 
electromagnetically induced transparency (EIT). EIT is a quantum interference
effect characterized by the presence of a frequency region with greatly 
reduced absorption accompanied by steep dispersion for a weak probe field 
propagating in a three-level atomic medium whose adjacent transition is 
driven by a strong coherent field \cite{eit_rev,pldpbook}. 

Optical lattices loaded with cold atoms under the EIT conditions offer
unique advantages over the previously studied schemes. First, the 
transparency bandwidth and the steep dispersion of the EIT resonance can 
be easily controlled by the corresponding driving field \cite{eit_rev}.
Second, for deep enough lattice potential, the Mott insulator regime 
can be reached \cite{mottOL}, in which each site will contain a single 
tightly localized atom. This will practically eliminate the coherence 
relaxation on the two-photon Raman transition caused by inter-atomic 
collisions and atomic time of flight, which result in a residual 
absorption of weak (quantum) fields and limit interaction times in 
most EIT experiments. Simultaneously, due to the tight localization of 
the atoms, the local atomic density is increased by two to three orders
of magnitude, as compared to the atomic density in the usual homogeneous
(gaseous) media. This, in turn, strongly enhances the modulation 
amplitude of the effective refractive index, resulting in a strong 
Bragg reflection of the probe field. The high-contrast periodic index
modulation is necessary for achieving 2D and 3D PBGs and strongly 
trapping the weak probe field. By contrast, in a uniform EIT medium 
\cite{swStEITPBG}, the dispersion modulation with an off-resonant 
standing wave field can only yield the index contrast up to a few percent,
which is sufficient for 1D Bragg reflection over extended propagation 
lengths amounting to thousands of periods, but not enough for strongly 
trapping the light in 2D or 3D with several hundred periods in each direction. 

The paper is organized as follows. In Sec.~\ref{sec:math}, the polarization
of the atomic medium for a weak probe field under the EIT conditions is 
rigorously derived. This is then used to obtain the 1D coupled-mode 
equations for the probe field. The solution of these equations and the 
resulting reflection, transmission and absorption spectra for the probe field
in the cold atomic medium are presented in Sec.~\ref{sec:rta}. A comparison 
with the case of a thermal atomic gas is made in  Sec.~\ref{sec:therm}. 
The conclusions are summarized in Sec.~\ref{sec:conc}.

\section{Mathematical Formalism}
\label{sec:math}

\begin{figure}[t]
\centerline{\includegraphics[width=6.5cm]{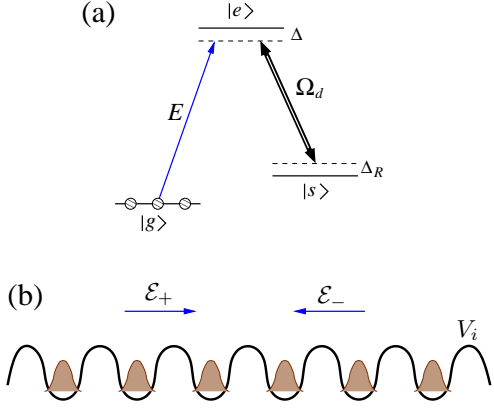}}
\caption{(a) Level scheme of cold atoms interacting with a weak 
probe field $E$ and strong driving field with Rabi frequency $\Om_d$.
(b) The atoms are trapped in a tight-binding optical lattice 
potential $V_i$.}
\label{as-ol}
\end{figure}

Consider the interaction of a weak probe field with an ensemble of 
cold bosonic (alkali) atoms with level configuration shown in 
Fig.~\ref{as-ol}(a). The probe field $E$ acts on the transition from 
the ground $\ket{g}$ to the excited $\ket{e}$ state. The transition 
$\ket{e} \lra \ket{s}$ is driven by a strong (near)resonant field 
with Rabi frequency $\Om_d$. The atoms in the lower meta-stable states
$\ket{g}$ and $\ket{s}$ are subject to external spatially periodic 
potential $V_{g,s}(\br)$ schematically shown in Fig.~\ref{as-ol}(b). 
The external potential $V_{e}$ for the excited state atoms is assumed 
negligibly small. All the atoms are initially prepared in the ground 
state $\ket{g}$. At sufficiently low temperatures, only the lowest 
Bloch band of the lattice potential is populated. In the limit of large 
lattice depth, this band reduces to the lowest energy levels of 
individual potential wells. Then, for a lattice filling factor 
$\rho \simeq 1$, and provided that tunneling rate of the atoms between 
the lattice sites is smaller than on-site interaction energy due to the
$s$-wave scattering, the Mott-insulator phase with single atom per 
lattice site will be attained \cite{OptLatRev,mottOL}.   

\subsection{Polarization of the Cold Atomic Medium}

The medium is described by three bosonic field operators $\psih_i(\br)$ 
representing atoms in the corresponding internal state $\ket{i}$ ($i=g,e,s$)
and possessing the standard bosonic commutation relations
$[\psih_i (\br), \psih_{i'}^{\dag} (\brp)] = \delta_{ii'}\,\delta(\br - \brp)$
and $[\psih_i (\br), \psih_{i'} (\brp)] 
= [\psih_i^{\dag} (\br), \psih_{i'}^{\dag} (\brp)] = 0$. 
In general, the Hamiltonian of the system has the form
\be
H = H_{A} + H_{AA} + H_{F} + H_{AF} \label{Hamtot}.
\ee 
Here $H_{A}$ is the atomic Hamiltonian 
\be
H_{A} = \sum_{i} \int d^3 r \, \psih_i^{\dag} (\br) 
\left[-\frac{\hbar^2}{2M} \nabla^2  + V_i (\br) + \hbar \om_i \right] 
\psih_i(\br) \label{Hamat},
\ee
with $M$ being the atomic mass and $\hbar \om_i$ the internal energy 
of the atoms in state $\ket{i}$. The term $H_{AA}$ describes the 
interatomic collisions, which, under the conditions of single atom per 
lattice site and negligible intersite tunneling assumed above, does not 
play a role \cite{OptLatRev,mottOL}. Next, if the probe field is assumed 
quantized, its Hamiltonian $H_{F}$ must be included in Eq.~(\ref{Hamtot}).
In the regime of a linear-response of the medium to a weak field studied 
here, the propagation dynamics of a quantized or a classical field is the
same and the probe field can be treated classically. Finally, $H_{AF}$ 
describes the interaction of the atoms with the probe 
$E(\br,t) \equiv \Ee(\br) e^{-i \om t}$ and driving
$E_d(\br,t) \equiv \Ee_d e^{i (\bk_d \cdot \br - \om_d t)}$ fields,
\bea
H_{AF} &=& \int d^3 r \, 
\psih_e^{\dag} (\br) \big[- \wp_{eg} E(\br,t) \big] \psih_g (\br) 
\nonumber \\ & & 
\int d^3 r \, 
\psih_e^{\dag} (\br) \big[- \wp_{es} E_d (\br,t) \big] \psih_s (\br)
+ \textrm{H. c.} ,
\eea
where $\wp_{ii'}$ is the dipole matrix element of the corresponding atomic
transition $\ket{i} \lra \ket{i'}$, and $\textrm{H. c.}$ stands for the 
Hermite conjugate. With the Hamiltonian (\ref{Hamtot}), the Heisenberg 
equations of motion for the atomic operators $\psih_i$ read
\besa{psihs}
\ddt \psih_g(\br) & = & -\frac{i}{\hbar} 
\left[-\frac{\hbar^2}{2M} \nabla^2 + V_g (\br) + \hbar \om_g \right] 
\psih_g(\br)
\nonumber \\ & &
+ i \frac{\wp_{ge}}{\hbar} \Ee^*(\br) e^{i \om t}  \psih_e(\br), \\
\ddt \psih_e(\br) & = & -\frac{i}{\hbar} 
\left[-\frac{\hbar^2}{2M} \nabla^2  + \hbar \om_e \right] 
\psih_e(\br)
\nonumber \\ & &
+ i \frac{\wp_{eg}}{\hbar} \Ee(\br) e^{-i \om t}\psih_g(\br) 
\nonumber \\ & &
+ i \Om_d \, e^{i (\bk_d \cdot \br - \om_d t)} \psih_s(\br), \qquad \\
\ddt \psih_s(\br) & = &  -\frac{i}{\hbar} 
\left[-\frac{\hbar^2}{2M} \nabla^2  + V_s (\br) + \hbar \om_s \right] 
\psih_s(\br)
\nonumber \\ & &
+ i \Om_d^* \, e^{-i (\bk_d \cdot \br - \om_d t)} \psih_e(\br), 
\eesa
where $\Om_d = \wp_{es}\Ee_d/\hbar$ is the drive Rabi frequency. 

The lattice potential is induced by far off-resonant standing-wave fields.
It has the form 
$V_{i}(\br) = \hbar \sum_{\xi = x,y,z} S_{i;\xi} \cos^2 (k_s \xi)$,
where the modulation (ac Stark shift) amplitude $S_{i;\xi}$ is 
proportional to the dynamic polarizability of the corresponding 
atomic state $\ket{i}$ ($i = g,s$) and the intensity of the 
standing-wave fields with the wave vector $k_s$. As stated above, 
in a deep OL potential, the atomic wavefunction is tightly 
localized at each lattice site $j$. The atomic field operators 
$\psih_i(\br)$ can then be expanded in terms of the real, 
normalized (Wannier) functions $w_i^{(j)}(\br) = w_i(\br - \br_j)$,
which in the vicinity of $\br \sim \br_j$ satisfy 
\be
\left[-\frac{\hbar^2}{2M} \nabla^2  + V_i (\br) \right] w_i^{(j)}(\br)
= \hbar \nu_i w_i^{(j)}(\br) ,
\ee
where $\hbar \nu_i$ is the energy of the lowest vibrational state 
of the nearly harmonic potential well, while $w_i^{(j)}(\br)$ 
can well be approximated by 3D Gaussian functions
\be
w_i^{(j)}(\br) = \left( \frac{1}{\pi \de r^2} \right)^{3/4} 
\exp \left[ - \frac{(\br - \br_j)^2}{2 \de r^2} \right] , \label{GaussWann}
\ee
centered around $\br_j$ with the width $\de r$. For tight localization
$\de r \ll \pi/ k_s$, the Wannier functions pertaining to different 
lattice sites have negligible overlap, 
$\int d^3 r \, w_i^{(j)}(\br) \, w_i^{(j')}(\br) = \de_{jj'}$.
The field operators for the lower atomic states can then be 
decomposed as $\psih_g(\br,t) = \sum_j w_g^{(j)}(\br) \, \gh_j(t) 
\, e^{-i (\om_g + \nu_g) t}$ and 
$\psih_s(\br,t) = \sum_j w_s^{(j)}(\br) \, \sh_j(t) 
\, e^{-i (\om_g + \nu_g) t} \, e^{- i (\om - \om_d) t}$, where
$\gh_j(t)$ and $\sh_j(t)$ are the slowly-varying in time annihilation 
operators for the bosonic atoms at site $j$. Since the atoms in the
excited state $\ket{e}$ are assumed free, $V_e \simeq 0$, the 
corresponding atomic field operator can be expanded as
$\psih_e(\br,t) = \sum_{\bk_l} u_{\bk_l}(\br) \, 
\eh_{\bk_l} (t) \, e^{-i (\om_g + \nu_g) t} \, e^{- i \om t}$,
where $u_{\bk_l}(\br) = (L)^{-3/2} e^{i \bk_l \cdot \br}$
are the plane waves within the quantization (medium) volume $L^3$, 
which satisfy
\bea
& & -\frac{\hbar^2}{2M} \nabla^2 u_{\bk_l}(\br)
= \hbar \nu_{k_l} u_{\bk_l}(\br) , \\
& & \int d^3 r \, u^*_{\bk_l}(\br) \, u_{\bk_l'}(\br) 
= \de_{\bk_l \bk_l'}, \quad 
\hbar \nu_{k_l} = \frac{\hbar^2 k_l^2}{2M} , \label{u-normcnd}
\eea
and $\eh_{\bk_l} (t)$ are the slowly-varying in time annihilation 
operators for the corresponding mode. 

Substituting the above decompositions of the atomic field operators
into Eqs.~(\ref{psihs}) yields
\besa{psiexp}
\sum_j w_g^{(j)}(\br) \, \dot{\gh}_j 
& = & -\ga_g \sum_j w_g^{(j)}(\br) \, \gh_j 
\nonumber \\ & &
+ i \frac{\wp_{ge}}{\hbar} \Ee^*(\br) \sum_{\bk_l} u_{\bk_l}(\br) \eh_{\bk_l}
+ \Fh_g, \label{atopg} \\
\sum_{\bk_l} u_{\bk_l}(\br) \, \dot{\eh}_{\bk_l}
& = & \sum_{\bk_l} (i \De - i \nu_{k_l} - \ga_e) 
u_{\bk_l}(\br) \, \eh_{\bk_l}
\nonumber \\ & &
+ i \frac{\wp_{eg}}{\hbar} \Ee(\br) \sum_j w_g^{(j)}(\br) \, \gh_j 
\nonumber \\ & &
+ i \Om_d \, e^{i \bk_d \cdot \br} \sum_j w_s^{(j)}(\br) \, \sh_j + \Fh_e, 
\label{atope} \\
\sum_j w_s^{(j)}(\br) \, \dot{\sh}_j  & = & 
(i \De_R -\ga_s) \sum_j w_s^{(j)}(\br) \, \sh_j 
\nonumber \\ & &
+ i \Om_d^* \, e^{-i \bk_d \cdot \br} \sum_{\bk_l}  
u_{\bk_l}(\br) \, \eh_{\bk_l} + \Fh_s , \label{atops} \quad
\eesa
where $\De \equiv \om + \om_g + \nu_g - \om_e$ 
and $\De_R \equiv \om + \om_g + \nu_g - \om_s - \nu_s - \om_d$ are, 
respectively, the one-photon and two-photon (Raman) detunings of 
the probe field [cf. Fig.~\ref{as-ol}(a)], while $\ga_i$ are the
phenomenological half-decay rates of the corresponding atomic states 
$\ket{i}$ ($i=g,e,s$). Note that $\ga_e \sim 10^7\:$s$^{-1}$ is large, 
determined by the short radiative lifetime of the excited atomic state,
while for isolated cold atoms trapped in a conservative OL potential,
$\ga_{g,s} \lesssim 1\:$s$^{-1}$ are negligibly small, due to the absence
of interatomic collisions and atomic free-flight or diffusion away for 
the interaction region. Finally, $\Fh_i$ are the $\de$-correlated noise
operators associated with the relaxation. These noise operators, required
to preserve the formal consistency of the equations for the atomic operators,
will not play a role in the following semiclassical treatment of the field 
propagation and will be dropped from now on.  

Under the weak probe approximation, the initial atomic population
in the ground state $\ket{g}$ remains practically undepleted and 
Eqs.~(\ref{psiexp}) can be solved to the lowest (first) order in 
$\Ee(\br)$. In the stationary regime, Eq.~(\ref{atops}) yields
\[
\sum_j w_s^{(j)}(\br) \, \sh_j 
= - i \frac{\Om_d^* \, e^{-i \bk_d \cdot \br} \sum_{\bk_l}  
u_{\bk_l}(\br) \, \eh_{\bk_l}}
{i \De_R -\ga_s} .
\]
Substituting this into Eq.~(\ref{atope}), multiplying all terms 
by $\int d^3 r \, u^*_{\bk_l}(\br)$ and using Eq.~(\ref{u-normcnd})
leads to
\be
\eh_{\bk_l} = - i \frac{\wp_{eg}}{\hbar} \,
\frac{\sum_j \int d^3 r \, u^*_{\bk_l}(\br) \, 
\Ee(\br) \, w_g^{(j)}(\br) \, \gh_j}
{i \De - \ga_e + \frac{|\Om_d|^2}{i \De_R -\ga_s} - i \nu_{k_l}} . 
\label{atopesol}
\ee
The integral in the above equation can be evaluated assuming that 
the localization width $\de r$ of $w_g^{(j)}(\br)$ is small compared
to the wavelength $\la = 2 \pi c/ \om $ of the probe field.
Therefore $\Ee(\br)$ changes little in the vicinity of $\br \simeq \br_j$
and can be taken constant in this region. The remaining integral
is then given by
\bea
\Ie^{(j)}_{\bk_l} &=& \int d^3 r \, u^*_{\bk_l}(\br) 
\, w_g^{(j)}(\br) \nonumber \\ 
&=& \left( \frac{1}{\pi \de r^2} \right)^{3/4} 
\big(\sqrt{2 \pi} \, \de r \big)^3
u^*_{\bk_l}(\br_j) e^{- \de r^2 \bk_l^2/2} . \label{Ikl}
\eea

The polarization $P(\br,t)= \Pe(\br) e^{-i \om t}$ of the atomic medium, 
induced by the applied probe field $E(\br,t)$, is given by the expectation 
value of the atomic dipole moment at position $\br$ and time $t$, 
$P(\br,t) = \big\langle \psih_g^{\dag}(\br,t) \, \wp_{ge} \, 
\psih_e(\br,t) \big\rangle$. With the above decompositions of the 
atomic field operators $\psih_{g,e}$ and using Eq.~(\ref{atopesol}),
the medium polarization $\Pe(\br)$ reads
\bea
\Pe(\br) &=& \wp_{ge} \Big\langle \sum_j w_g^{(j)}(\br) \, \gh_j^{\dag} 
\sum_{\bk_l} u_{\bk_l}(\br) \, \eh_{\bk_l} \Big\rangle
\nonumber \\ 
&=& -i \frac{|\wp_{eg}|^2}{\hbar} \sum_{j,j'} w_g^{(j)}(\br) \Ee(\br_{j'}) 
\big\langle \gh_j^{\dag} \gh_{j'} \big\rangle \nonumber \\ & & \times 
\sum_{\bk_l} \frac{u_{\bk_l}(\br) \, \Ie^{(j')}_{\bk_l}}
{i \De - \ga_e + \frac{|\Om_d|^2}{i \De_R -\ga_s} - i \nu_{k_l}} . \qquad
\eea
Here the sum over the plane-wave modes $\bk_l$ 
should be replaced by an integral according to 
$\sum_{\bk_l} \to (L/2 \pi)^3 \int d^3 k_l$ \cite{pldpbook}, which, 
upon substituting $\Ie^{(j')}_{\bk_l}$ from Eq.~(\ref{Ikl}), reads
\bea
& &\left( \frac{L}{2 \pi} \right)^3 \left( \frac{1}{\pi \de r^2} \right)^{3/4} 
\big(\sqrt{2 \pi} \, \de r \big)^3 \nonumber \\ 
& & \times \int d^3 k_l 
\frac{u_{\bk_l}(\br) \, u^*_{\bk_l}(\br_{j'}) \, 
e^{- \de r^2 \bk_l^2/2}} 
{i \De - \ga_e + \frac{|\Om_d|^2}{i \De_R -\ga_s} - i \nu_{k_l}} .
\label{Intklmodes}
\eea
It is easy to see that if the atomic localization width is not too small, 
$\de r^2 \gg \de r^2_{\textrm{min}} = \hbar/M \ga_e$ \cite{comment},
$\nu_{k_l} = \frac{\hbar k_l^2}{2M}$ in the denominator of the above equation
contributes little to the integral and can therefore be dropped. The 
integral is then easily evaluated, and Eq.~(\ref{Intklmodes}) reduces to
\[
\frac{w_g^{(j')}(\br)}{i \De - \ga_e + \frac{|\Om_d|^2}{i \De_R -\ga_s}} .
\]
Since the Wannier functions $w_g^{(j)}(\br)$ and $w_g^{(j')}(\br)$ 
pertaining to different lattice sites $j \neq j'$
have negligible overlap, the final expression for the medium 
polarization takes the form
\be
\Pe(\br) = i \frac{|\wp_{eg}|^2}{\hbar} \frac{\varrho(\br) \, \Ee(\br) }
{\ga_e - i \De + \frac{|\Om_d|^2}{\ga_s - i \De_R}} , \label{Pr-fin}
\ee
where $\varrho(\br) = \sum_j |w_g^{(j)}(\br)|^2 
\big\langle \gh_j^{\dag} \gh_j \big\rangle$ is the spatially-periodic 
density of the medium. According to the assumption above, the 
occupation number of each lattice site is 
$\big\langle \gh_j^{\dag} \gh_j \big\rangle = 1$. 
Clearly, the spectral response of the medium coincides with that of the 
conventional EIT with homogeneous atomic ensemble \cite{eit_rev,pldpbook},
but is spatially modulated by the periodic atomic density, which leads 
to profound consequences for the probe transmission and reflection, 
as discussed below. 

\subsection{Coupled-Mode Equations for the Probe Field}

The propagation of electromagnetic field in the medium
is governed by the Maxwell wave equation
\be
\left[ \nabla^2  - \mu_0 \eps_0 \frac{\partial^2}{\partial t^2} \right]
E(\br,t) = \mu_0 \frac{\partial^2}{\partial t^2} P(\br,t) . \label{Maxwell}
\ee
In free space, $P = 0$, its general solution for a monochromatic field
of frequency $\om$ has the form 
$E(\br,t) = \sum_{\bk} \Ee_{\bk} e^{i (\bk \cdot \br - \om t)}$,
where $\Ee_{\bk}$ is the amplitude of the field mode with wave vector 
$\bk$ satisfying $|\bk| = \om/c$, with $c = ( \mu_0 \eps_0)^{-1/2}$
being the speed of light in vacuum. 

Consider a stationary propagation of the probe field in the periodic 
atomic medium whose spatial and spectral properties are characterized
by Eq.~(\ref{Pr-fin}). In particular, assume that a circularly ($\sigma^-$)
polarized probe field propagates along the $\hat{\mathbf{z}}$ axis 
(taken to be the quantization axis), while a linearly ($\pi$) polarized 
driving field, propagating along the $\hat{\mathbf{x}}$ axis 
($\bk_d \bot \hat{\mathbf{z}}$ ), uniformly irradiates the whole atomic 
sample. Following the approach of \cite{yariv}, the probe field can be 
expanded in terms of the free-space normal modes as 
$E(z,t) = \sum_{k} \Ee_{k}(z) e^{i (k z - \om t)}$, where
$\Ee_{k}(z)$ are now slowly varying in space amplitudes of the 
corresponding field modes $k = \pm \om/c$. The medium polarization 
can be expressed as $P(z,t) = \eps_0 \chi(\om; z) E(z,t)$, where
$\chi(\om; z)$ is the linear susceptibility, whose $z$-dependence
is due to the atomic density $\varrho(z)$. This 1D density is obtained by
averaging $\varrho(\br)$ over the transverse to $\hat{\mathbf{z}}$ directions
of the cubic lattice with primitive vector $\La = \pi / k_s$,
\be
\varrho (z) \equiv \frac{1}{\La^2} \int_{-\La/2}^{\La/2} \!\! dx 
\int_{-\La/2}^{\La/2} \!\! d y \, \varrho(\br) . 
\ee
Since $\varrho (z) = \varrho(z + \La)$ is a periodic function, 
it can be expanded in a Fourier series 
\be
\varrho (z) = \sum_l \varrho_l \, e^{i l g z} , 
\quad g = \frac{2 \pi}{\La} = 2 k_s , \\
\ee
with the expansion coefficients
\bea
\varrho_l &=& \frac{1}{\La} \int_{-\La/2}^{\La/2} \!\! d z \, \varrho(z) \, 
e^{- i l g z} \nonumber \\ 
&=& \frac{\big\langle \gh_j^{\dag} \gh_j \big\rangle}{\La^3} 
\int_{-\La/2}^{\La/2} \!\! d z \, |w_g(z)|^2 \, e^{- i l g z} ,
\eea
where, according to Eq.~(\ref{GaussWann}), 
\be
|w_g(z)|^2 = \frac{1}{\sqrt{\pi} \de r}  
\exp \left[ - \frac{z^2}{\de r^2} \right]
\ee
is a normalized Gaussian function. Thus, the zeroth Fourier component
$\varrho_0 = \big\langle \gh_j^{\dag} \gh_j \big\rangle/\La^3$ 
($\de r \ll \La$) is the spatially averaged (bulk) atomic density. 
Higher order Fourier components can be expressed as 
$\varrho_l = \varrho_0 \ka_l$, where
\be
\ka_l \equiv \int_{-\La/2}^{\La/2} \!\! d z \, |w_g(z)|^2 \, e^{- i l g z} .
\ee
Using Eq.~(\ref{Pr-fin}), the medium polarization then reads
\be
P(z,t) = \eps_0 \, \chi(\om) \sum_l \sum_{k'}
\ka_l \, \Ee_{k'}(z) \, e^{i (l g + k')z}  \, e^{- i \om t} , \label{Pr-exp}
\ee 
where
\be
\chi(\om) \equiv \chi_{\textrm{EIT}}(\om) 
= 2 \, \frac{c}{\om} \, \sigma_0 \varrho_0 \, 
\frac{i \ga_e} {\ga_e - i \De + \frac{|\Om_d|^2}{\ga_s - i \De_R}} , 
\label{chiEIT}
\ee
is the usual EIT susceptibility of an atomic medium with uniform 
density $\varrho_0$ and resonant absorption cross-section 
$\sigma_0 = |\wp_{eg}|^2 \om_{eg}/(2 \eps_0 c \hbar \ga_e)$ \cite{pldpbook}. 
By definition, the resonant amplitude absorption coefficient in the 
two-level atomic medium ($\Om_d = 0$) is $a_0 = \sigma_0 \varrho_0$.  

Using Eq.~(\ref{Pr-exp}), Maxwell's equation (\ref{Maxwell}) yields
\bea
& & 2 i \sum_k k \left[\frac{d}{d z} \Ee_{k}(z) \right] e^{i k z} \nonumber \\
& & \qquad = - \frac{\om^2}{c^2} \chi(\om) \sum_l \sum_{k'}
\ka_l \, \Ee_{k'}(z) \, e^{i (l g + k')z} ,  
\eea  
which, upon expanding the sums over $k, k' = \pm \om/c$, 
leads to the following coupled equations for the amplitudes 
$\Ee_{\pm } \equiv \Ee_{\pm \om/c}$ of the forward ($+$) 
and backward ($-$) propagating modes,  
\besa{cmodSeqs}
\frac{d}{d z} \Ee_{+} &=& i \alpha(\om)
\sum_l \kappa_l \big[\Ee_{+} \, e^{i l g z} 
+ \Ee_{-} \, e^{i (l g - 2 \om/c) z}   \big] , \\
\frac{d}{d z} \Ee_{-} &=& - i \alpha(\om)
\sum_l \kappa_l \big[\Ee_{+} \, e^{i (l g + 2 \om/c) z}  
+ \Ee_{-} \, e^{i l g z} \big] , \qquad
\eesa
where
\be
\alpha(\om) \equiv \frac{\om}{c} \frac{\chi(\om)}{2} 
= \frac{i a_0 \ga_e} {\ga_e - i \De + \frac{|\Om_d|^2}{\ga_s - i \De_R}} .
\label{alphaEIT}
\ee
Note that $\kappa_l$ decrease as $|l|$ increase. Thus, apart from 
$\kappa_0 = 1$, $\kappa_{\pm 1}$ are the largest coefficients given by
($\de r \ll \La$)
\be
\ka_{\pm 1} \simeq  
 \frac{1}{\sqrt{\pi} \de r} \int_{-\infty}^{\infty} \!\! d z \,
e^{- z^2/ \de r^2} \cos (2 \pi z/\La) .  
\ee
Assuming, therefore, that $\om/c \sim k_s$ and keeping only the
terms satisfying the longitudinal phase matching condition (i.e.,
slowly oscillating in space), Eqs.~(\ref{cmodSeqs}) reduce to
\besa{cmodeqs}
\frac{d}{d z} \Ee_{+} &=& i \alpha(\om) \Ee_{+} 
+ i \alpha(\om) \kappa_1 \Ee_{-} \, e^{- 2 i (\om/c - k_s) z} , \\
\frac{d}{d z} \Ee_{-} &=& - i \alpha(\om) \Ee_{-} 
- i \alpha(\om) \kappa_{-1} \Ee_{+} \, e^{2 i (\om/c - k_s) z} . \qquad
\eesa 
In these equations, the first term describes the usual EIT absorption
and dispersion of the probe field propagating in the atomic medium,
while the second term is responsible for the coupling of the forward
and backward propagating modes mediated by the medium periodicity.

\section{Reflection, Transmission and Absorption of the Probe}
\label{sec:rta}

The coupled mode equations (\ref{cmodeqs}) fully determine the 
optical properties of the system under consideration. The solution 
of Eqs.~(\ref{cmodeqs}) for the boundary value problem defined through 
$\Ee_{+}(0) = \Ee_{\textrm{in}}$ and $\Ee_{-}(L) = 0$ is given by
\besa{amplsol}
\Ee_{+}(z) &=&  \Ee_{\textrm{in}} \, e^{- i (\om/c - k_s) z} \nonumber \\
& & \times 
\frac{s \cosh \big[s (L-z)\big] - i \de \beta \sinh \big[s (L-z)\big]}
{s \cosh \big[ s L \big] - i \de \beta \sinh \big[s L \big]} , \qquad \\ 
\Ee_{-}(z) &=&  \Ee_{\textrm{in}} \, e^{i (\om/c - k_s) z} \nonumber \\
& & \times 
\frac{i \alpha(\om) \, \ka_{-1} \sinh \big[s (L-z)\big]}
{s \cosh \big[ s L \big] - i \de \beta \sinh \big[s L \big]} ,
\eesa
where the coefficients
\beann
\de \beta &=& \alpha(\om) + \frac{\om}{c} - k_s , \\
s &=& \sqrt{ \alpha^2(\om) \ka_1 \ka_{-1} - \de \beta ^2} ,
\eeann
quantify, respectively, the phase mismatch and the coupling between 
the forward and backward propagating modes.

\begin{figure}[t]
\centerline{\includegraphics[width=8.7cm]{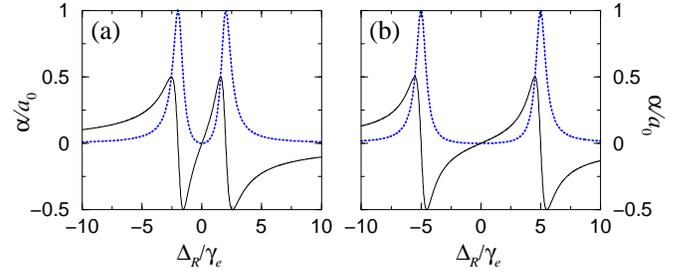}}
\caption{Absorption (blue dotted lines) and dispersion (black solid lines) 
spectra for the probe field in a uniform EIT atomic medium. 
The Rabi frequency of the resonant driving field is 
(a)~$\Om_d = 2 \ga_e$, and (b)~$\Om_d = 5 \ga_e$.}
\label{EITalpha}
\end{figure}

According to the Bloch theorem \cite{pc-books,yariv}, the general 
solution for the electromagnetic field propagation in the periodic
medium can be cast as $\Ee(z) = e^{i K z} \Ee_K(z)$, where $K$ is a 
propagation constant known as the Bloch wave vector, while 
$\Ee_K(z) = \Ee_K(z + \La)$ is a spatially periodic function. 
In a purely dispersive (non-absorbing) medium, the real and imaginary 
parts of $K$ describe, respectively, the spatial phase and the attenuation
(due to Bragg reflection) of the field upon propagation. It follows 
from Eqs.~(\ref{amplsol}) that the corresponding propagation constant 
for the monochromatic probe field of frequency $\om$ is given by
\bea
K &=& k_s + i s \nonumber \\
&=& k_s + i \sqrt{\left[ \alpha (\om) \ka_1 \right]^2 
- \left[\alpha (\om) + \frac{\om}{c} - k_s \right]^2} . \qquad \label{disprel}
\eea 
In the present situation, the periodic atomic medium exhibits both
strong dispersion and absorption characterized by the EIT polarizability
$\alpha (\om)$ (see Fig.~\ref{EITalpha}). Therefore, in the dispersion
relation of Eq.~(\ref{disprel}), both Bragg reflection and medium 
absorption would contribute to the imaginary part of $K$. A more 
quantitative and experimentally accessible characterization of the 
optical properties of the system is provided by the reflection $R$, 
transmission $T$ and absorption $A$ coefficients for the probe field, 
which are defined through
\besa{TRA}
R (\om) & \equiv & \left| \frac{\Ee_{-}(0)}{\Ee_{+}(0)} \right|^2 , \\
T (\om) & \equiv & \left| \frac{\Ee_{+}(L)}{\Ee_{+}(0)} \right|^2 , \\
A (\om) & = & 1 - \big[ R (\om) + T (\om) \big] .
\eesa

The absorption and dispersion spectra for the probe field in a uniform 
atomic medium subject to a resonant drive $\om_d = \om _{es} - \nu_s$ 
is shown in Fig.~\ref{EITalpha}. The strong driving field with Rabi 
frequency $\Om_d \gtrsim \ga_e$ splits the familiar Lorentzian absorption
line into the Autler-Townes doublet with the two peaks separated by 
$2 \Om_d$, while at the line center the medium becomes transparent to 
the probe field. This EIT is accompanied by a steep normal dispersion, 
resulting in the greatly reduced group velocity \cite{eit_rev,pldpbook}
\be
v_g = \frac{c}{1+c \frac{\partial}{\partial \om} 
\textrm{Re} \, \alpha(\om) } \simeq \frac{|\Om_d|^2}{a_0 \ga_e} \ll c . 
\ee

The principal aim here is to explore the possibility of a PBG for the 
probe field in the frequency region of the reduced absorption. Assuming 
resonant driving field, so that $\De_R = \De$, in the vicinity of EIT 
resonance $|\De| < \Om_d$, the polarizability of Eq.~(\ref{alphaEIT})
reduces to $\alpha(\om) \simeq \De /v_g$ and the dispersion relation
(\ref{disprel}) can be approximated by
\be 
K - k_s \simeq i \sqrt{\left[ \frac{\De}{v_g} \ka_1 \right]^2 
- \left[\frac{\De}{v_g} - \frac{\de_s}{c} \right]^2} , 
\ee
where $\de_s \equiv k_s c - \om_{eg}$. The Bloch wave vector $K$ has an 
imaginary part when the term under the square-root of this equation is 
positive. The corresponding range of frequencies of the probe field is
\be
\frac{\de_s v_g}{(1 \pm \ka_1) c} < \De <  \frac{\de_s v_g}{(1 \mp \ka_1) c} ,
\quad \mbox{for} \quad \de_s \gtrless 0 . \label{PBGedges}
\ee 
Thus, a PBG of width 
$\de \om_{\textrm{gap}} = |\de_s| \frac{v_g}{c} \frac{2 \ka_1}{1 - \ka_1^2}$,
centered at $\De_{\textrm{gap}} = \de_s \frac{v_g}{c} \frac{1}{1 - \ka_1^2}$,  
is formed on the positive (for $\de_s > 0$) or negative (for $\de_s < 0$)
side of the EIT resonance. The peak value of the imaginary part
of $K$, which determines the maximal reflectivity at 
$\De_{\textrm{gap}}$, is given by 
\be
\max (\textrm{Im} \, K) = \frac{|\de_s|}{c} 
\frac{\ka_1}{\sqrt{1 - \ka_1^2}} . \label{maxImK}
\ee
These conclusions are justified upon requiring that the PBG lies within 
the EIT window $\de \om_{\textrm{tw}} = |\Om_d|^2/(\ga_e \sqrt{2 a_0 L})$
\cite{pldpbook,eitwtr} where the absorption is small. This 
leads to the following condition on the lattice wavevector $k_s$,  
\be
|k_s - \om_{eg}/c| \equiv \frac{|\de_s|}{c} 
\lesssim (1 - \ka_1^2) \sqrt{\frac{a_0}{2 L}} , \label{dscond}
\ee
with which
$\max (\textrm{Im} \, K) \lesssim \ka_1 \sqrt{a_0 (1 - \ka_1^2) /2L}$.
With the system parameters listed in \cite{comment}, and for an 
atomic medium of length $L \simeq 200\:\mu$m (500 lattice periods $\La$),
or optical depth $2 a_0 L =100$, the above condition is satisfied for 
$|k_s - \om_{eg}/c| \simeq  4.5 \times 10^{3}\:$rad/m or 
$|\de_s| /2 \pi \simeq 2.16 \times 10^{11}\:$s$^{-1}$. Physically, 
such a large difference between the lattice wave vector $k_s$ and 
that of the probe field $k \simeq \om_{eg}/c$ stems from the need 
to satisfy the phase matching condition $\de \beta \simeq 0$
[while $\alpha(\om) \ka_1 \neq 0$] in strongly dispersive  EIT medium.     

\begin{figure}[t]
\centerline{\includegraphics[width=8.7cm]{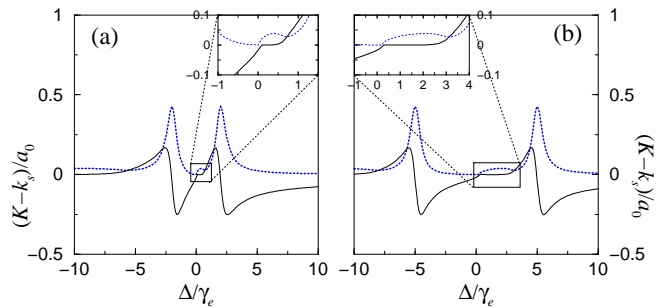}}
\caption{Imaginary (blue dotted lines) and real (black solid lined) parts 
of $K$ as a function of probe detuning $\De$ (resonant drive, $\De = \De_R$), 
for $\ka_1 = 0.9$ ($\de r \simeq \La/10$), $\de_s = 7.2 \times 10^4 \, \ga_e$,
and (a)~$\Om_d = 2 \ga_e$ and (b)~$\Om_d = 5 \ga_e$. Insets magnify
the important frequency regions within the EIT window.}
\label{BlochK}
\end{figure}

The complete dispersion relation of Eq.~(\ref{disprel}), with 
$\ka_1 \simeq 0.9$ obtained for $\de r \simeq \La/10$ \cite{comment},
is plotted in Fig.~\ref{BlochK} for the two values of $\Om_d$ used
in Fig.~\ref{EITalpha}. As noted above, the imaginary part of $K$
describes exponential attenuation of the forward propagating 
probe field due to the back reflection and medium absorption.
Within the transparency window, $|\De| \lesssim \de \om_{\textrm{tw}}$,
the absorption is much smaller than the spatially-periodic dispersion,
and the bulges of $\textrm{Im} \, K$ in the insets of Fig.~\ref{BlochK} 
signify the appearance of the PBG. The band edges given by 
Eq.~(\ref{PBGedges}) can be tuned by changing the drive field intensity,
since $v_g \propto |\Om_d|^2$.

\begin{figure}[b]
\centerline{\includegraphics[width=8.7cm]{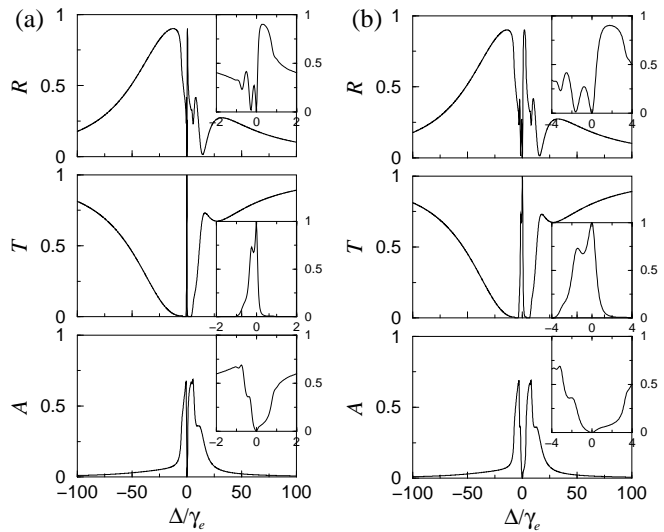}}
\caption{Reflection $R$, transmission $T$ and absorption $A$ spectra
for the probe field in a cold EIT atomic medium loaded into an OL. 
The medium length $L =500 \La \simeq 200\:\mu$m, 
or optical depth $2 a_0 L = 100$, and all the other parameters
are same as in Fig.~\ref{BlochK}(a) and (b), respectively. 
Insets magnify the important frequency regions within 
the EIT window.}
\label{RTAolEIT}
\end{figure}

Figure~\ref{RTAolEIT} shows the reflection, transmission and absorption
spectra for the probe field for the two values of $\Om_d$ used in 
Figs.~\ref{EITalpha} and \ref{BlochK}. The peak reflectivity for the 
PBG within the EIT window is about 90\%, limited mainly by absorption.
Choosing smaller values of $\de_s$, thereby moving the PBG closer to 
the EIT line center, and taking simultaneously longer medium, to 
compensate for smaller values of $\max (\textrm{Im} \, K)$ as per 
Eq.~(\ref{maxImK}), will yield even larger values of the maximum 
reflection (up to 98\%), at the expense of narrower PBG.  

\begin{figure}[t]
\centerline{\includegraphics[width=7.0cm]{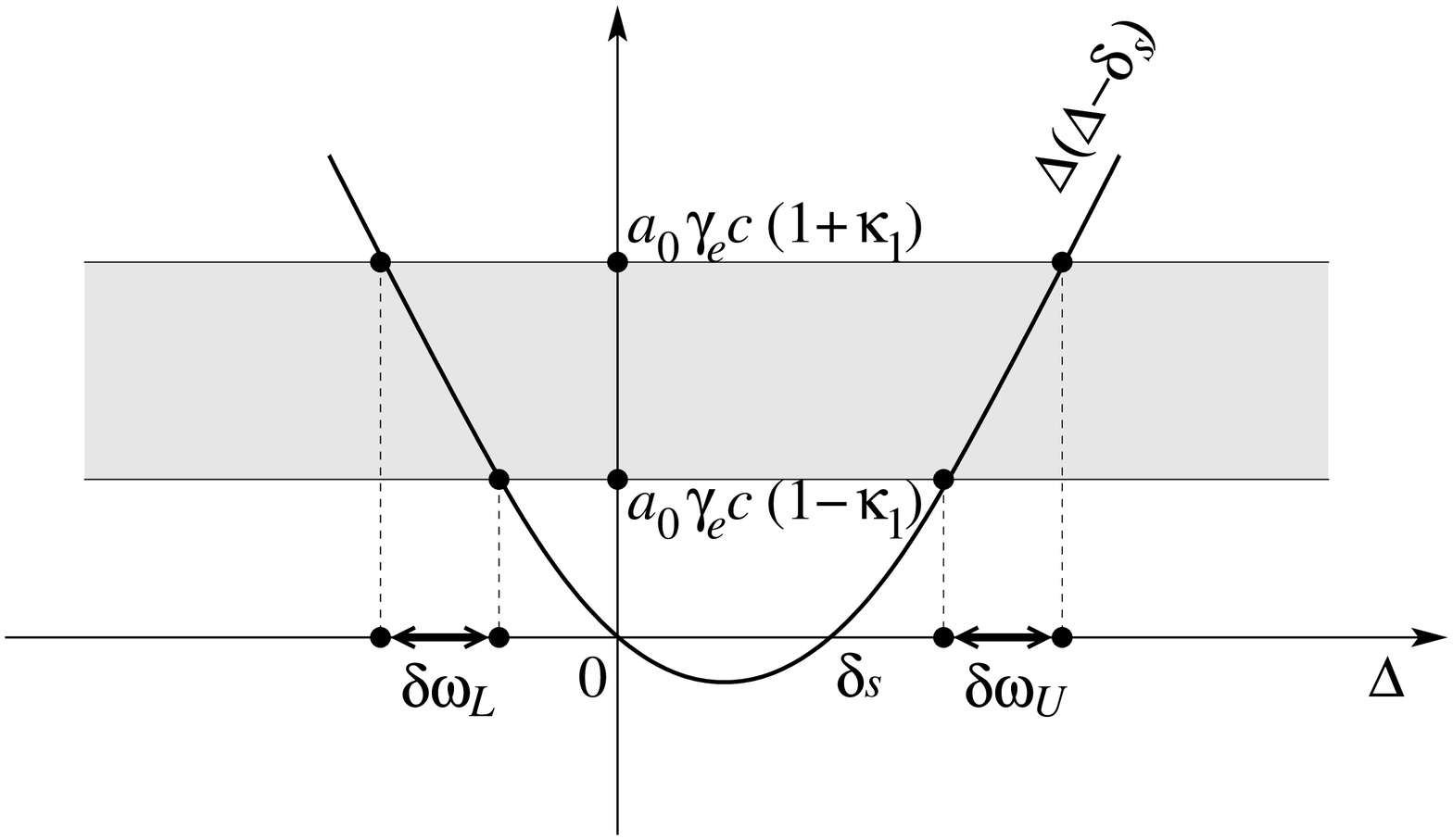}}
\caption{Diagram illustrating the appearance of the lower $\de \om_L$
and upper $\de \om_U$ PBGs far away from the EIT resonance, where the 
atoms behave as off-resonant two-level systems with negligible absorption.}
\label{tlaPBG}
\end{figure}

In the spectral region far away from the atomic line center and EIT
features, there is another broad PBG seen in the reflection spectrum of
Fig.~\ref{RTAolEIT}. Its existence was predicted and studied in 
\cite{Phillips} for the limiting case of $\ka_1 \to 1$, corresponding 
to the atomic localization width $\de r \to 0$. Briefly, in the frequency 
regions $|\De| \gg \ga_e, \Om_d$, the polarizability of Eq.~(\ref{alphaEIT})
can be approximated as $\alpha(\om) \simeq - a_0 \ga_e/\De$, which describes
the response of off-resonant two-level atoms, and the dispersion relation
reads
\be 
K - k_s \simeq i \sqrt{\left[ \frac{a_0 \ga_e}{\De} \ka_1 \right]^2 
- \left[\frac{a_0 \ga_e}{\De} + \frac{\de_s -\De}{c} \right]^2} .
\ee
The Bloch wave vector acquires imaginary part when
$a_0 \ga_e c (1 - \ka_1) < \De (\De - \de_s) < a_0 \ga_e c (1 + \ka_1)$,
which is graphically illustrated in Fig.~\ref{tlaPBG}. Two PBGs appear,
one on the red $\De < 0$ and the other on the blue $\De > 0$ sides of 
the atomic resonance. The corresponding ranges of frequencies for the
lower ($L$) and upper ($U$) gaps are
\besa{tlabgedgs}
\hlf \de_s - D_{+} < \De_L < \hlf \de_s - D_{-} , \\
\hlf \de_s + D_{-} < \De_U < \hlf \de_s + D_{+} ,
\eesa
where $D_{\pm} \equiv \sqrt{(\de_s/2)^2 + a_0 \ga_e c (1 \pm \ka_1)}$,
and the widths of the gaps are $\de \om_{L,U} = D_{+} - D_{-}$. 
When $(\de_s/2)^2 \gg a_0 \ga_e c$, which for $\de_s$ chosen as in
Eq.~(\ref{dscond}) is certainly satisfied, $D_{\pm}$ can be expanded as
$D_{\pm} \simeq |\de_s|/2 + a_0 \ga_e c (1 \pm \ka_1)/|\de_s|$.
For $\de_s > 0$, the lower gap $- a_0 \ga_e c (1 + \ka_1)/|\de_s| 
< \De_L < - a_0 \ga_e c (1 - \ka_1)/|\de_s|$ is closer to the atomic 
resonance, and the corresponding reflection is seen in Fig.~\ref{RTAolEIT};
the upper gap is beyond the range of frequencies spanned in that figure.
Note finally that the reflection, transmission and absorption spectra 
for $\de_s > 0$ and $\de_s < 0$ are related by mirror symmetry about $\De =0$.

\section{Thermal Atomic Medium}
\label{sec:therm}

For the sake of comparison, it is instructive to consider the 
case of a thermal atomic gas under the otherwise similar conditions. 
When the thermal energy $k_{\textrm{B}} T$ is much larger than 
the depth of the OL potential $\hbar S_i$, the atoms move freely 
and the atomic density $\varrho$ is uniform \cite{doppler}. 
However, the off-resonant standing wave field with wave-vector $k_s$ 
results in a spatially-periodic ac Stark-shift $\hbar S_i \cos^2 (k_s z)$
of the lower atomic levels $\ket{i}$ ($i=g,s$). The resulting susceptibility
\be
\chi(\om ; z) = 2 \, \frac{c}{\om} \,  
\frac{i a_0 \ga_e} {\ga_e - i \De(z) + \frac{|\Om_d|^2}{\ga_s - i \De_R(z)}} , 
\label{chiEITsw}
\ee  
is a periodic function of $z$, since the corresponding one- and 
two-photon detunings $\De(z) = \De' + \frac{1}{2} S_g \cos (2 k_s z)$ 
and $\De_R(z) = \De_R' + \frac{1}{2} S_{gs} \cos (2 k_s z)  $ are ac 
Stark modulated. Here $\De' = \om - \om_{eg} + \frac{1}{2} S_g$ and
$\De_R' = \om - \om_d + \om_{gs} + \frac{1}{2} S_{gs}$ are the mean 
detunings, while $S_{gs} = S_g - S_s$ is the difference of the 
ac Stark modulation amplitudes for levels $\ket{g}$ and $\ket{s}$. 
The decay rate $\ga_s$ describes the Raman coherence relaxation,
which is now affected by the atomic motion and collisions; its typical 
value is in the range of $\ga_s \sim 10^3 -10^4\:$s$^{-1}$ \cite{eit_rev},
which is still much smaller than $\ga_e$ but not negligible. The effect
of the spatially-periodic modulation of the Raman detuning $\De_R(z)$ 
is to periodically shift the EIT spectrum for the probe field
(see Fig.~\ref{EITalpha}), which under appropriate conditions discussed 
below can result in a PBG, as was shown in \cite{swStEITPBG}. 

\begin{figure}[t]
\centerline{\includegraphics[width=8.7cm]{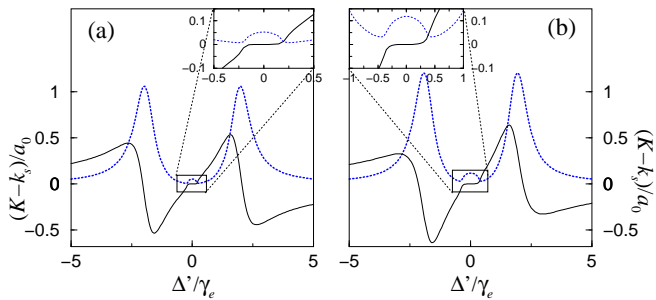}}
\caption{Imaginary (blue dotted lines) and real (black solid lines) parts of 
$K$ as a function of probe detuning $\De'$ (resonant drive, $\De' = \De_R'$), 
for $\Om_d = 2 \ga_e$, $\ga_s = 10^{-4} \ga_e$ and $\de_s =0$. 
The ac Stark shift amplitudes are (a) $\frac{1}{4} S_{gs} = 0.2 \ga_e$, 
and (b)~$\frac{1}{4} S_{gs} = 0.4 \ga_e$. Insets magnify the 
important frequency regions within the EIT window.}
\label{BlochKsw}
\end{figure}

Substituting the susceptibility of Eq.~(\ref{chiEITsw}) into the Maxwell 
equation (\ref{Maxwell}) with $P(z,t) = \eps_0 \chi(\om; z) E(z,t)$ and 
$E(z,t) = \sum_{k} \Ee_{k}(z) e^{i (k z - \om t)}$, under the EIT 
conditions $|\Om_d|^2 \gg \ga_e \ga_s, (\frac{1}{4}S_{g,s})^2$ and the 
longitudinal phase matching $k_s \simeq \om/c$, the following coupled 
mode equations for the forward and backward propagating modes 
$k= \pm \om/c$ of the probe field are obtained,
\besa{cmodeqssw}
\frac{d}{d z} \Ee_{+} &=& i \alpha'(\om) \Ee_{+} 
+ i \eta(\om) \Ee_{-} \, e^{- 2 i (\om/c - k_s) z} , \\
\frac{d}{d z} \Ee_{-} &=& - i \alpha'(\om) \Ee_{-} 
- i \eta(\om) \Ee_{+} \, e^{2 i (\om/c - k_s) z} , \qquad
\eesa 
where
\besa{alphaeta}
\alpha'(\om) & =& \frac{a_0 \ga_e} 
{\big[|\Om_d|^2 - \De' \De_R'\big]^2 + \ga_e^2 \De_R^{\prime 2} } 
\nonumber \\ & & \times
\Big\{ \De_R' (|\Om_d|^2 -\De' \De_R')  
\nonumber \\ & & \qquad 
+ i \big[ |\Om_d|^2 \ga_s + \ga_e ( \De_R^{\prime 2} 
+ \mbox{$\frac{1}{8}$} S_{gs}^2) \big] \Big\} , \\
\eta(\om) & =& \frac{a_0 \ga_e} 
{\big[|\Om_d|^2 - \De' \De_R'\big]^2 + \ga_e^2 \De_R^{\prime 2} } 
\nonumber \\ & & \times \frac{1}{4}
\Big\{ |\Om_d|^2 S_{gs} - \De_R'( 2 \De' S_{gs} + \De_R' S_g)  
\nonumber \\ & & \qquad 
+ 2 i \ga_e \De_R' S_{gs}) \Big\}. 
\eesa
The solution of Eqs.~(\ref{cmodeqssw}) is given by Eqs.~(\ref{amplsol}),
with
\[
\de \beta = \alpha'(\om) + \frac{\om}{c} - k_s , \quad
s = \sqrt{ \eta^2(\om) - \de \beta ^2} .
\]
The corresponding dispersion relation $K = k_s + i s$, with a resonant
driving field $\om_d = \om_{es} - \frac{1}{2} S_s$ ($\De_R' = \De'$)
and $k_s \simeq \om_{eg} /c$, is shown in Fig.~\ref{BlochKsw},
where nonzero values of $\textrm{Im} \, K$ within the EIT window  
signify the appearance of a PBG for the probe field.

In the vicinity of EIT resonance $|\De'| < \Om_d$ (assuming 
resonant drive, $\De' = \De_R'$), Eqs.~(\ref{alphaeta}) reduce to
$\alpha'(\om) \simeq \De' /v_g$ and $\eta(\om) \simeq S_{gs} / (4 v_g)$,
and the dispersion relation can be approximated as
\be 
K - k_s \simeq i \sqrt{\left[ \frac{S_{gs}}{4 v_g} \right]^2 
- \left[\frac{\De'}{v_g} - \frac{\de_s}{c} \right]^2} . 
\ee
For $\de_s =0$, a PBG for the probe field with frequencies in the 
range $|\De'| <  \frac{1}{4} |S_{gs}|$ is formed 
($\de \om_{\textrm{gap}} = \frac{1}{2} |S_{gs}|$). The peak value of 
$\textrm{Im} \, K$ attained at the gap center $\De'_{\textrm{gap}} = 0$ 
is $\max (\textrm{Im} \, K) = |S_{gs}|/(4 v_g)$. Note that a PBG 
for the probe field exists only when $S_{gs} \equiv S_g - S_s \neq 0$, 
i.e., the ac Stark shifts $S_g$ and $S_s$ of the lower atomic levels 
$\ket{g}$ and $\ket{s}$ induced by an off-resonant standing wave field 
should be different. It is clear that in order to minimize absorption, the 
ac Stark modulation of the Raman resonance should be accommodated with
the EIT window, $\frac{1}{2} |S_{gs}| \lesssim \de \om_{\textrm{tw}}$ 
\cite{swStEITPBG}. This, in turn, restricts the peak reflectivity 
of the medium to $\max (\textrm{Im} \, K) \lesssim \sqrt{a_0 /8L}$.   
 
\begin{figure}[t]
\centerline{\includegraphics[width=8.7cm]{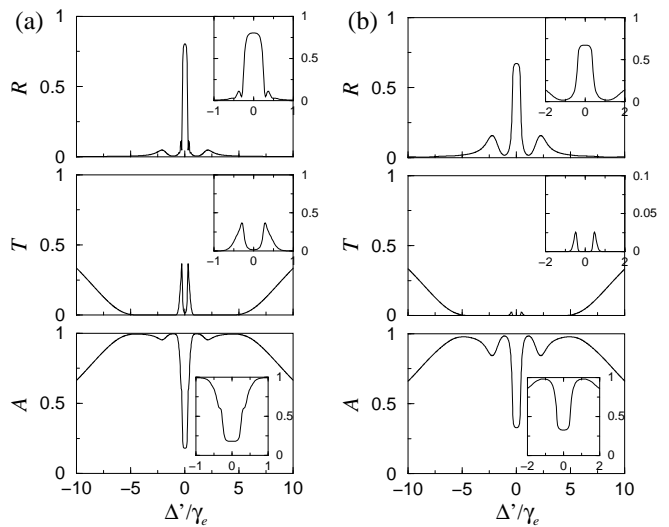}}
\caption{Reflection $R$, transmission $T$ and absorption $A$ spectra
for the probe field in a thermal EIT atomic medium subject to spatially
periodic ac Stark modulation of the Raman resonance. The medium length 
is $L \simeq 200\:\mu$m (optical depth $2 a_0 L = 100$), and all the 
other parameters are same as in Fig.~\ref{BlochKsw}(a) and (b), respectively. 
Insets magnify the important frequency regions within the EIT window.}
\label{RTAswEIT}
\end{figure}

Figure~\ref{RTAswEIT} shows the reflection, transmission and 
absorption spectra for the probe field for the two values of 
$S_{gs}$ used in Fig.~\ref{BlochKsw}. In Fig.~\ref{RTAswEIT}(a) with 
$\frac{1}{4} S_{gs} = 0.2 \ga_e$, the peak of the reflection 
coefficient $R$ at $\De' = 0$ is about 80\% and the absorption $A$
is 18\%. Even though increasing $S_{gs}$ results in broader band gap 
and larger values of $\textrm{Im} \, K$, as seen in Fig.~\ref{BlochKsw}(b)
with $\frac{1}{4} S_{gs} = 0.4 \ga_e$, the corresponding reflection 
coefficient of Fig.~\ref{RTAswEIT}(b) is smaller, $R \simeq 67\%$, 
due to the increased absorption $A \simeq 33\%$. Physically, this 
can be understood by recalling (see Fig.~\ref{EITalpha} and 
\cite{eit_rev,pldpbook}) that in the vicinity of EIT resonance, 
the dispersion (which determines $\textrm{Im} \, K$) scales linearly
with detuning $\De_R$ (i.e., with $S_{gs}$), while absorption scales 
quadratically with $S_{gs}$. Therefore, by choosing smaller values of 
the ac Stark modulation amplitudes $S_{gs}$, and taking simultaneously
longer medium, to compensate for reduced $\textrm{Im} \, K$, will 
yield larger values of the reflection coefficient at the center of 
the PBG which will however shrink in width.

\section{Conclusions and Outlook}
\label{sec:conc}

Equation (\ref{chiEIT}) or (\ref{alphaEIT}) characterize the spectral 
properties of electromagnetically induced transparency (EIT) in 
an atomic medium, as illustrated in Fig.~\ref{EITalpha}. 
Looking at that figure, one could come up with three possible ways 
for achieving spatially-periodic index modulation 
$\de n (\om; z) \simeq \frac{1}{2} \textrm{Re} \, \chi(\om ; z)$ 
resulting in a photonic band gap (PBG) in the vicinity of EIT resonance 
$|\De_R| < \Om_d,\ga_e$, where the susceptibility is approximately given by
\[
\chi(\om ; z) \simeq 2 \, \frac{c}{\om} \,  
\frac{\sigma_0 \varrho \ga_e }{|\Om_d|^2} \, \De_R.
\] 
(i) To periodically modulate the medium density $\varrho = \varrho(z)$
by trapping cold atoms in an optical lattice, as proposed and studied here.
(ii) To periodically shift the EIT spectrum $\De_R = \De_R(z)$ in a
thermal atomic ensemble by using spatially periodic ac Stark shift of 
the Raman transition induced by off-resonant standing-wave field, 
as proposed in \cite{swStEITPBG} and reviewed in Sec.~\ref{sec:therm} above.
(iii) To use a standing-wave drive field $\Om_d = \Om_d (z)$ resulting 
in a spatially periodic modulation of the bandwidth of the EIT and the 
associated with it dispersion slope, as studied in \cite{swDrEITPBG}.  

The novel scheme proposed here possesses a number of advantageous
properties. These include: (a) Greatly reduced coherence relaxation
on the two-photon Raman transition due to the elimination of atomic 
diffusion and collisions causing probe absorption. (b) Simultaneous 
enhancement of the refractive index modulation and the resulting Bragg
reflection due to the tight localization of the atoms. (c) Tunability of 
the position and the width of the PBG within and beyond the EIT resonance.
These properties can be employed for achieving very efficient nonlinear 
interactions between weak quantum fields and realization of deterministic
quantum logic with single photons, as described in \cite{XPM-PBG}.

Finally, it would be interesting to explore the possibility of achieving
a 2D PBGs for the probe field in the medium of cold atoms trapped in an 
optical lattice. The ultimate goal would be to explore structures
with appropriately engineered defects which may allow for strong 
localization and waveguiding of light. Such defects can easily be 
implemented in the laboratory experiments by simply focusing a resonant 
laser onto the desired lattice sites, which will release (evaporate) 
the corresponding atoms from the trap. In contrast, the microfabrication
of desired defects in solid-state photonic crystal structures 
\cite{pc-books} involves complicated and time consuming growth and 
lithographic techniques which are often not easily reconfigurable and 
reproducible. Thus tunable photonic band gaps in optical lattices may 
serve to experimentally test various proposed structures in a simple, 
quickly reproducible way.

\begin{acknowledgments}
I gratefully acknowledge helpful discussions with M.~Fleischhauer 
and G.~Kurizki. This work was supported in part by the EC 
Marie-Curie Research Training Network EMALI, and in part by 
the Alexander von Humboldt Foundation during my stay at the 
Technische Universit\"at Kaiserslautern, where this project 
was initiated.
\end{acknowledgments}


\begin{thebibliography}{10}
\bibitem{doppler}
It is assumed that temperature $T$ is still low enough 
for the Doppler broadening to be insignificant, 
$\om/c \, \sqrt{2 k_{\textrm{B}} T/M} \ll \ga_e, \Om_d$, which 
for the parameters of \cite{comment} requires $T \lesssim 1\:$mK.
\end{thebibliography}
\end{document}